\documentclass[11pt, a4paper]{article}
\pdfoutput=1 

\usepackage{jheppub} 

\usepackage[T1]{fontenc} 

\usepackage[utf8]{inputenc}
\usepackage{epstopdf}
\usepackage[footnotesize]{caption}

\def \be {\begin{equation}}
\def \ee {\end{equation}}
\def \bea {\begin{eqnarray}}
\def \eea {\end{eqnarray}}
\def \nn {\nonumber}

\def \rr {\raise.35ex\hbox{\small $\prime$}\kern-.17em{\mbox{\large $\imath$}}}

\def \dels {\partial\kern-.6em /\kern.1em}
\def \As {{A\kern-.5em / \kern.5em}}
\def \Ds {D\kern-.7em / \kern.5em}

\def \ks {k\kern-.5em /}
\def \ls {l\kern-.5em /}

\title{One-Loop $\beta$ Function of the Double Sigma Model with Constant Background}

\author[a, 1]{Chen-Te Ma \note{Corresponding author.}}

\affiliation[a]{Department of Physics, Center for Theoretical Sciences and
Center for Advanced Study in Theoretical Sciences, 
National Taiwan University, Taipei 10617, Taiwan,
R.O.C.}

\emailAdd{yefgst@gmail.com}

\abstract{The double sigma model with the strong constraints is equivalent to a classical theory of the normal sigma model with one on-shell self-duality relation. The one-form gauge field comes from the boundary term. It is the same as the normal sigma model. The gauge symmetries under the strong constraints are the diffeomorphism and one-form gauge transformation in the double sigma model. These gauge symmetries are also the same as the Dirac-Born-Infeld (DBI) theory. The main task of this work is to compute one-loop $\beta$ function to obtain the low energy effective theory of the double sigma model. We implement the self-duality relation in the action to perform the one-loop calculation. At last, we obtain the DBI theory. We also rewrite this theory in terms of the generalized metric and scalar dilaton, and define the generalized scalar curvature and tensor from the equations of motion.}

\arxivnumber{1412.1919}

\begin{document} 
\maketitle
\flushbottom

\section{Introduction}
\label{1}
The most important problem of physics is to answer how to unify all fundamental theories. One intelligent way is duality. It is the main idea of the M-theory. We can use the T-duality and S-duality to unify all ten dimensional superstring theories. If we combine the T-duality and S-duality, it is the so-called U-duality. We expect that the U-duality is a symmetry of the M-theory. However, the M-theory is still mysterious at the current stage. The main problem is that we do not completely understand our tool, duality. The S-duality is an equivalence between strong and weak coupling constant. The familiar example is invariance of the Maxwell's equations by exchanging electric and magnetic fields. Because this duality should be a non-perturbative duality, it is difficult to study from the perturbative way explicitly. The other one duality, T-duality, is an equivalence between radius and inverse radius on a compact torus. This duality is equivalent to exchanging momentum and winding modes in closed string theory, or the Dirichlet and Neumann boundary conditions in open string theory. However, there is one remaining serious problem that we cannot solve it. This problem is called T-fold problem. This problem is found in closed string theory. It is mainly due to the fact that the T-duality is not a well-defined transition function as gauge transformation or diffeomorphism in the presence of non-zero flux. For the low energy massless closed string field theory with the $H$-flux \cite{Zwiebach:1992ie, Saadi:1989tb}, we can perform the T-duality on one direction to let the $H$-flux to become the $f$-flux. At this step, it is still well-defined. If we perform the second T-duality, we will get the $Q$-flux. The problem occurs because fields cannot be described as single valued. For the third T-duality, it gives a more serious problem, we do not know how to perform this third T-duality because we lose isometry. However, we expect that the $R$-flux can be found by the T-duality. In this T-fold problem, we meet two problems, the first problem is how to define our fields with single values in a new geometry and the second problem is how to extend our T-duality definition to obtain the $R$-flux. Solving the T-fold problem should give us a new perspective to our M-theoretical frame work. It should lead us to understand a new supergravity or superstring theory.  

The dynamics of the superstring or M-theory is hard to obtain by the first principle directly. One way is to study low energy effective theory. At the level of field theory, we can understand symmetry principle and dynamics. String theory can be described by a two dimensional sigma model. From low energy effective theory, we can understand what kind of low energy effective theory can be realized on the target space. It leads us to understand corresponding gauge symmetry on target space. Low energy physics inspires us to study non-local field theories beyond the standard model and normal particle physics. Non-local theories help us to develop techniques to study dynamics of field theory. The development not only gives us a new way to study partition functions or amplitudes on field theories, but also the conceptual aspects of the M-theory. However, the M-theory is still mysterious, it is difficult to write a consistent Lagrangian to describe it now. Nevertheless, the low energy effective theory of the M2-M5 system has already been constructed. One consistent single M5 system is the Nambu-Poisson M5 (NP M5) \cite{Ho:2008nn, Ho:2012dn}. The way of the construction is analogous to the stack of the D$p$-brane in the B-field background to obtain the D($p+2$)-brane theory. The Nambu-Poisson M5-brane theory can be stacked from the C-field background of the multiple M2-brane.  It gives us a single M5-brane in the large constant C-field background (Only three spatial dimensions  are not zero.) The coupling of this single M5-brane is the inverse C-field background. The role of the Nambu-Poisson M5-brane theory is similar to the non-commutative D-brane theory. The symmetry of the Nambu-Poisson M5-brane theory is the volume preserving diffeomorphism (VPD) governed by the Nambu-Poisson bracket which satisfies the fundamental identity. The consistency of a single M5-brane theory is to perform the direct dimensional reduction to find the non-commutative D4-brane in a constant NS-NS B-field background. This consistent check is already shown. The problem is that they only obtained the Poisson bracket, not the deformed version. It also implies that the Nambu-Poisson M5-brane is just a truncated M5-brane theory. Even for a truncated M5-brane theory, it is still interesting to study new theories by the dualities. By performing a double dimensional reduction on the Nambu-Poisson M5-brane, we can obtain the non-commutative D4-brane in a large constant R-R C-field background. It can be generalized to the D$p$-brane based on the gauge symmetry,  covariant field strength, rotational symmetry of the scalar field and duality rules. This D$p$-brane is built on the non-commutative space in a large constant R-R ($p-1$)-form background. The NS-NS D$p$-brane, R-R D$p$--brane, and Nambu-Poisson M5-brane are well-defined low energy effective theories under the decoupling limit. Especially, the NS-NS D3-brane and R-R D3-brane theories are also consistent with the electric-magnetic duality. It shows that the Nambu-Poisson M5-brane theory has consistency on the T-duality and S-duality \cite{Ho:2013opa, Ho:2013paa, Ma:2012hc}. These studies are also interesting to see relations between the background and brackets. The symmetry of the ($p-1$)-form background theory can be described by the ($p-1$)-form bracket exactly. The most important direction of this single M5-brane is the way of deformation. The hint of the deformation can be found from the direction of the S-duality because we also have the same problem on the D-brane theory. The way is to use all orders of the non-commutative NS-NS D3-brane to find the deformed non-commutative R-R D3-brane by the electric-magnetic duality. This study gives us a new product to write this R-R D3-brane theory. It should indicate how to write the full-order low energy effective theory consistent with dualities. These Nambu-Poisson M5-brane studies not only answer problems of the M-theory, but also other difficult problems of field theories, e.g., the electric-magnetic duality of the non-abelian gauge theory in four dimensions. Since the $U(1)$ non-commutative gauge theory is similar to the non-abelian gauge theory, we can point out that the electric-magnetic duality of the non-commutative gauge theory inspires us to solve the electric-magnetic duality of the non-abelian gauge theory. This way can be understood from the string duality. It is a good example to show that string duality not only concerns unification, but also new structure of field theories.

The Nambu-Poisson M5-brane theory is built on the non-commutative space by the stack of the multiple M2-brane theory. From the perspective of non-commutative geometry, we should be possible to build the M5-brane from the equivalence of commutative and non-commutative gauge theories or from the Seiberg-Witten map. For the DBI theory (a string ending on a $p$-brane), we need to change from the closed to open string parameters. We can find this redefinition from the Poisson-Sigma model. For a higher form field, we can consider the Nambu-Sigma model \cite{Jurco:2012yv, Schupp:2012nq}, which is classically equivalent to a $p$-brane theory. From the Nambu-Sigma model, we can change variables to consider a non-commutative theory. Starting from the equivalence of non-commutative and commutative gauge theories, we can find the form of field theories without many degrees of freedom. This theory is called generalized DBI theory, which describes a $q$-brane ending on a $p$-brane. This generalized DBI theory can reduce to the DBI theory when $q$=1. In the case of a two brane ending on a five brane, we can find the same form for the M5-brane action up to the second order perturbatively (derivative expansion) \cite{Jurco:2012yv, Schupp:2012nq}. From dimensional reduction on this special case, we find that the two brane ending on the five brane becomes a one brane ending on a four brane \cite{Ho:2014una}. Even though this calculation is not a general consideration, this consistency on the DBI-form of the M5-brane still gives a strong support. This approach is to offer a new generalized metric, which gives a new structure of non-commutative theory. It gives another way to obtain the same theory as the generalized DBI \cite{Jurco:2013upa, Jurco:2014sda}. It not only shows that the equivalence of non-commutative and commutative gauge theories for an arbitrary form field can be described by a new generalized metric, but also shows that this equivalence is strongly restricted to our action. On the other hand, supergravity interpretation of the generalized DBI theory should have supersymmetric extension. The related supersymmetric extension has already been done in \cite{Lee:2010ey}. The Nambu structure of the $p$-brane theory can be found manifestly on the formulation of taking off the square root \cite{Park:2008qe}. It gives a consistent understanding with the Nambu-Poisson M5-brane. It may indicate that the $p$-brane theory has some relations with the M-theory.

The Nambu-Poisson M5 and the generalized DBI theories are still defined on the local geometry so they do not really strike the main problem of the T-fold. Although their constructions have already provided insight to the M-theory, they may not solve the T-fold problem or the global geometry. Now we have a way to find a new geometry to understand string theory. The new geometry is called \lq\lq stringy geometry\rq\rq \cite{Hohm:2011dv, Hohm:2010xe}. This way is to double coordinates. It embeds the T-duality rule in the $O(D, D)$ group. This type of theory is called double field theory \cite{Hull:2009mi, Hohm:2010jy, Lee:2013hma, Tseytlin:1990va, Tseytlin:1990nb, Copland:2011wx, Berman:2007yf, Berman:2007xn, Avramis:2009xi, Duff:1989tf, Berkeley:2014nza, Siegel:1993xq, Siegel:1993th, Siegel:1993bj,  Hull:2009zb, Hohm:2010pp}. Now we  have a well-defined theory to describe the massless closed string field theory with the strong constraints. Although the strong constraints are equivalent to removing half additional coordinates, this theory is not a new theory without considering the T-duality. Double field theory is a way to provide the extension of the T-duality or solve the T-fold problem \cite{Aldazabal:2011nj, Andriot:2012an}. The T-fold problem is that the local geometry is  unable to find a theory with the $R$-flux by using the T-duality. If we want to define a theory with the $R$-flux, we need to go beyond the original supergravity and T-duality. Double field theory is built on the doubled space. The gauge transformation is governed by the Courant bracket. On this doubled space, we can perform the T-duality three times to find the $R$-flux in the massless closed string field theory. It implies that we can use this doubled space to know how to perform the T-duality for non-isometric case. Double field theory extends the supergravity from local to global geometry. The so-called non-geometric flux ($Q$- and $R$-flux) can be understood from a geometric way. The understanding of the non-geometric flux has important influence on brane theory. The sources of the exotic brane theory are non-geometric fluxes. The exotic brane theory can be shown by performing the T-duality two times on the Neveu-Schwarz five-brane (NS5-brane). This exotic brane is called the $5_2^2$-brane theory. The background of the exotic brane is no longer single-valued. It implies that global description is needed. However, worldvolume theory for the $5_2^2$-brane theory is constructed from the NS5-brane theory by performing the T-duality two times \cite{Kimura:2014upa}. The exotic brane theory should play an important role on the extension of our understanding for the M5-brane because of the NS5-brane can be uplifted to the M5-brane theory. Although we have new concepts with the strong constraints in double field theory, we still want to relax the constraints. Because the strong constraints equivalently imply that solutions will be annihilated by the constraints. We want to relax the constraints to get more solutions that will not be annihilated. It is a very important study, but the closed property of the generalized Lie derivative makes this problem be a very hard task. The approaches can be seen in \cite{Geissbuhler:2013uka, Ma:2014ala}. For more extension of the original theory, we need to consider $\alpha^{\prime}$ corrections. We already formulate the theory in the language of double field theory \cite{Hohm:2013jaa} at the first step. Some good reviews of double field theory are in \cite{Hohm:2013bwa, Aldazabal:2013sca, Berman:2013eva}.

Double field theory is a formulation for the ten dimensional supergravity. The extension from the T-duality to U-duality or from ten dimensions to eleven dimensions, we need to consider exceptional field theory \cite{Hohm:2013vpa, Hohm:2013uia, Hohm:2014fxa}. The low energy limit of the M-theory is the eleven dimensional supergravity. The symmetry of the eleven dimensional supergravity should be the exceptional Lie group. For the analogue consideration of double field theory, we need to embed the exceptional Lie group into a bigger space. From the theoretical point of view of the M-theory, the manifest symmetry becomes important to give us the insight to know properties of the M-theory although we just know the low energy level. The first difficulty of this task is the $E_{8(8)}$ case, it does not have closed algebra. However, this problem is already solved by sacrificing some Lorentz gauge freedoms. The exceptional field theory helps us to show the U-fold problem as a double field theory. It also needs a constraint as double field theory. The current stage of relaxing constraint does not have too much progress. With the strong constraint of the exceptional field theory, we can obtain the exceptional generalized geometry. It provides intuition to realize the eleven dimensional supergravity to inspire exceptional field theory from a different way \cite{Berman:2010is}.

Double field theory extends string theory from local to global geometry. For a self-consistent double field theory, we need to extend our understanding of closed string theory to open string theory. Otherwise, we cannot write full string theory in terms of double field theory. Then double field theory may not be a fully consistent understanding of string theory. The first proposal of double field theory for open string is to double coordinates with two types of boundary conditions and also introduces projectors to satisfy the suitable boundary conditions \cite{Hull:2004in}. For the projectors, it does not give a full understanding until \cite{Albertsson:2008gq}. This paper extends the idea of the projectors to show a consistent boundary condition. However, their discussion only considers the background without the one-form gauge field. For the first study of the one-form gauge field, they put the normal boundary term after they introduce the self-duality relation at the off-shell level. They cannot consistently obtain the DBI action from the one-loop $\beta$ function \cite{Albertsson:2011ux}. From the generalized geometry \cite{Gualtieri:2003dx, Hitchin:2004ut, Cavalcanti:2011wu}, they used the Courant bracket to understand the properties of the D-brane \cite{Hatsuda:2012uk, Asakawa:2012px}. Especially for \cite{Asakawa:2012px}, they construct the gauge transformation based on the language of generalized geometry. It inspires \cite{Ma:2014kia} to find the gauge transformation of the open string theory in the language of double field theory. From the gauge transformation (governed by the $F$-bracket), the double sigma model is also proposed. The main difference is that the double sigma model does not use projectors to satisfy the boundary conditions, but they have the classical equivalence with the normal sigma model without modifying the self-duality relation. Since this double sigma model only puts the normal boundary term and the self-duality relation does not have modification with the one-form gauge field, the one-loop $\beta$ function can be performed in this double sigma model. Quantum fluctuation from string theory  inspires  the higher derivative gravity model \cite{Zwiebach:1985uq} so a calculable sigma model is undoubtedly necessary. The $R$-flux can be found from the Courant bracket without an action in \cite{Asakawa:2014kua}. The suitable action for the D-brane theory is proposed in \cite{Ma:2014kia}. It should give a consistent $R$-flux as \cite{Asakawa:2014kua}.

We implement the self-duality relation at the off-shell level with the strong constraints. Then we use the action to perform the one-loop $\beta$ function to obtain a consistent DBI theory. We rewrite this low energy effective theory in terms of the generalized metric and scalar dilaton. We also use the equations of motion to define the generalized Ricci scalar curvature and tensor. 

The plan of this paper is to first review the double sigma model in Sec. \ref{2}.
Then we calculate the one-loop $\beta$ function and obtain the low energy effective action in Sec. \ref{3}. We also discuss the generalized metric formulation, and show the generalized Ricci scalar curvature and tensor in Sec. \ref{4}.
Finally, we conclude and summarize in Sec. \ref{5}.

\section{Review of the Double Sigma Model}
\label{2}
We review the double sigma model in this section. We first show the notation and set up. Then we write the gauge transformation of the double sigma model. In the end of this section, we show classical equivalence between the double and normal sigma model. 

\subsection{Notation and Set Up}
Our theory is defined on the doubled space. The normal coordinates are associated with the Neumann boundary condition and the other coordinates (transverse coordinates) are associated with the Dirichlet boundary condition. The field components are the metric field ($g_{mn}$), antisymmetric background field ($B_{mn}$), scalar dilaton ($d$) and one-form gauge field ($A_m$). In this theory, we need two constraints (strong constraints) to guarantee the gauge invariance. The constraints are
\bea
\partial_m\tilde{\partial}^m(\mbox{field})=0,
\qquad
\partial_m\tilde{\partial}^m(AB)=0,
\eea
where
\bea
\partial_m=\frac{\partial}{\partial x^m},
\qquad
\tilde{\partial}^m=\frac{\partial}{\partial\tilde{x}_m}.
\eea
The index $m=0, 1\cdots, D-1$ (We denote the non-doubled target index from $m$ to $z$.).
If we only consider the first constraint, the conventional name is called weak constraint. Imposing the weak constraint leads to
\bea
\partial_m\tilde{\partial}^m\delta(\mbox{field})\neq 0,
\eea
where $\delta$ is the gauge transformation. For a consistent gauge invariance, we need to impose the strong constraints to annihilate non-gauge invariant parts. We can rewrite the weak constraint as
\bea
\partial^A\partial_A(\mbox{field})=0,
\eea
where $\partial_A$ is defined by
\be
{\partial_A }\equiv
 \begin{pmatrix} \,\tilde{\partial}^m \, \\[0.6ex] {\partial_m } \end{pmatrix}
\ee
and $\partial^A=\eta^{AB}\partial_C$. The index, $A=0, 1\cdots, 2D-1$, is a doubled dimensional target index (We represent them from $A$ to $K$.). We use $\eta$ to raise and lower the indices for $O(D, D)$ tensor
\be
 h= \begin{pmatrix} a& b \\ c& d \end{pmatrix}
\, ,
\qquad   h^t \eta h = \eta, \qquad\eta= \begin{pmatrix} 0& I \\ I& 0 \end{pmatrix},
\ee
 where $a$, $b$, $c$ and $d$ are $D\times D$ matrices.
We can also define $X^A$ from the combination of the normal and dual coordinates by
\be
{X^A}\equiv
 \begin{pmatrix} \,\tilde x_m\, \\[0.6ex] {x^m} \end{pmatrix}.
\ee

\subsection{$C$- and $D$-Bracket}
We introduce the generalized Lie derivative, $C$-bracket and $D$-bracket \cite{ Hohm:2010pp, Hull:2009zb}. 
The gauge transformation of the background independence field (${\cal E}$) and scalar dilaton are
\bea
\label{gauge tran_BI}
\delta{\cal E}_{mn}&\equiv&\delta(g+B)_{mn}={\cal D}_m\tilde{\xi}_n-\bar{{\cal D}}_n\tilde{\xi}_m+\xi^p\partial_p{\cal E}_{mn}+{\cal D}_m\xi^p{\cal E}_{pn}+\bar{{\cal D}}_{n}\xi^p{\cal E}_{mp},
\nn\\
\delta d&=&-\frac{1}{2}\partial_p\xi^p+\xi^p\partial_p d,
\eea
where
\bea
e^{-2d}=\sqrt{-\det{g}}e^{-2\phi},
\eea
\bea
{\cal D}_m=\partial_m-{\cal E}_{mn}\tilde{\partial}^n, \qquad \bar{{\cal D}}_m=\partial_m+{\cal E}_{nm}\tilde{\partial}^n,
\eea
 and $\phi$ is the dilaton. Then we define the symmetric $O(D, D)$ element (${\cal H}_{AB}$, generalized metric),
   \be
  {\cal H}~ \equiv ~ {\cal H}^{\bullet\,\bullet}  \,,
  \ee
\be
  {\cal H} \ = \
  \begin{pmatrix}    g-Bg^{-1}B & Bg^{-1}\\[0.5ex]
  -g^{-1}B & g^{-1}\end{pmatrix}\, .
 \ee
This matrix is defined by
 \be
 {\cal H}\,\eta \,{\cal H}=\eta\;.
  \ee
The inverse of ${\cal H}$ can be obtained by
\be
{\cal H}^{-1} =  \eta {\cal H} \eta\,,
\ee
\be
  {\cal H}^{-1}~ \equiv ~ {\cal H}_{\bullet\,\bullet}  \,\ = \ \left({\cal H}^{AB}\right)^{-1} \ = \
  \begin{pmatrix}    g^{-1} & -g^{-1}B\\[0.5ex]
  Bg^{-1} & g-Bg^{-1}B\end{pmatrix}\;.
  \ee
  The gauge transformation of ${\cal H}^{AB}$ is
\bea
\delta_{\xi} {\cal H}^{AB}=\xi^C\partial_C{\cal H}^{AB}+(\partial^A\xi_C-\partial_C\xi^A){\cal H}^{CB}+(\partial^B\xi_C-\partial_C\xi^B){\cal H}^{AC},
\eea
where
\be
{\xi^A}\equiv
 \begin{pmatrix} \,\tilde{\xi}_m\, \\[0.6ex] {\xi^m} \end{pmatrix}\equiv \begin{pmatrix} \,\Lambda_m\, \\[0.6ex] {\epsilon^m} \end{pmatrix}.
\ee
The generalized Lie derivative can be defined from the gauge transformation
\bea
\hat{\cal L}_{\xi}{\cal H}^{AB}\equiv\delta_{\xi}{\cal H}^{AB},
\eea
which satisfies the Leibniz rule. The special property of the generalized Lie derivative is the acting on the constant metric ($\eta$) \emph{is} zero, but the ordinary Lie derivative is not.
The gauge algebra is closed under the strong constraints
\bea
[\hat{\cal L}_{\xi_1}, \hat{\cal L}_{\xi_2}]=\hat{\cal L}_{[\xi_1, \xi_2]_C},
\eea
where the $C$-bracket is defined by
\bea
[\xi_1, \xi_2]_C^A=\xi_1^C\partial_C\xi_2^A-\xi_2^C\partial_C\xi_1^A-\frac{1}{2}\eta^{AC}\eta_{DE}\xi_1^D\partial_C\xi_2^E+\frac{1}{2}\eta^{AC}\eta_{DE}\xi_2^D\partial_C\xi_1^E.
\eea
The $D$-bracket for the generalized vector is defined by
\bea
[A, B]_D\equiv\hat{\cal L}_A B.
\eea
At the end of the $C$- and $D$-bracket, we assume that all parameters are independent of $\tilde{x}$ on the $C$-bracket. Then it reduces to the Courant bracket \cite{Hull:2009zb}.
\bea
\lbrack\xi_1, \xi_2\rbrack_C^m&=&\xi_1^p\partial_p\xi_2^m-\xi_2^p\partial_p\xi_1^m=({\cal L}_{\xi_1}\xi_2)^m\equiv([\xi_1, \xi_2])^m,
\nn\\
\lbrack\xi_1, \xi_2\rbrack_{Cm}&=&\xi^p_1\partial_p\tilde{\xi}_{2m}-\xi^p_2\partial_p\tilde{\xi}_{1m}
-\frac{1}{2}(\xi_1^p\partial_m\tilde{\xi}_{2p}-\tilde{\xi}_{2p}\partial_m\xi^p_1)
+\frac{1}{2}(\xi_2^p\partial_m\tilde{\xi}_{1p}-\tilde{\xi}_{1p}\partial_m\xi^p_2)
\nn\\
&=&
\xi^p_1\partial_p\tilde{\xi}_{2m}-\xi^p_2\partial_p\tilde{\xi}_{1m}
+(\partial_m\xi_1^p)\tilde{\xi}_{2p}-\frac{1}{2}\partial_m(\xi_1^p\tilde{\xi}_{2p})
-(\partial_m\xi_2^p)\tilde{\xi}_{1p}+\frac{1}{2}\partial_m(\xi_2^p\tilde{\xi}_{1p})
\nn\\
&=&\bigg({\cal L}_{\xi_1}\tilde{\xi}_2-\frac{1}{2}d(i_{\xi_1}\tilde{\xi}_2)\bigg)_m
-\bigg({\cal L}_{\xi_2}\tilde{\xi}_1-\frac{1}{2}d(i_{\xi_2}\tilde{\xi}_1)\bigg)_m.
\eea
It shows the Courant bracket
\bea
[A+\alpha, B+\beta]_{\mbox{Cour}}=[A, B]+{\cal L}_A\beta-{\cal L}_B\alpha-\frac{1}{2}d(i_A\beta-i_B\alpha),
\eea
where $A$, $B$ are vectors, and $\alpha$, $\beta$ are one-form. We also obtain the Dorfman bracket \cite{Gualtieri:2003dx} from the $D$-bracket.
\bea
[A+\alpha, B+\beta]_{\mbox{Dor}}=[A, B]+{\cal L}_A\beta-i_Bd\alpha.
\eea
For the consistent notation, we denote the Dorfman bracket in a different way instead of the conventional way $(A+\alpha)\circ(B+\beta)$. The $D$-bracket has the Jacobi identity
\bea
[A, [B, C]_D]_D=[[A, B]_D, C]_D+ [B, [A, C]_D]_D,
\eea
but it is not antisymmetric. For the $C$-bracket, it does not satisfy the Jacobi identity, but it is antisymmetric. In other words, the $C$- and $D$-bracket are \emph{not} Lie brackets.

\subsection{$F$-Bracket}
We discuss the $F$-bracket \cite{Ma:2014kia} in this section.
The gauge transformation of the gauge field is
\bea
\delta A_m&=&\Lambda_m+{\cal L}_{\epsilon}A_m.
\eea 
Then we calculate $\lbrack\delta_1,\delta_2\rbrack A_m=-\delta^{\prime}A_m$. 
\bea
\epsilon^{\prime m}&=&\epsilon_1^n\partial_n\epsilon_2^m-\epsilon_2^n\partial_n\epsilon_1^m,
\nn\\
\Lambda^{\prime}_m&=&\epsilon_1^{n}\partial_{n}\Lambda_{2m}
+(\partial_m\epsilon_1^n)\Lambda_{2n}-\epsilon_2^{n}\partial_{n}\Lambda_{1m}
-(\partial_m\epsilon_2^n)\Lambda_{1n}
\nn\\
&=&{\cal L}_{\epsilon_1}\Lambda_{2m}-{\cal L}_{\epsilon_2}\Lambda_{1m}.
\eea
We define the $F$-bracket from this closed algebra.
\bea
\lbrack\xi_1,\xi_2\rbrack_F^A=\bigg(\xi_1^D\partial_D\xi_2^A-\xi_2^D\partial_D\xi_1^A\bigg)
-\frac{1}{2}\bigg(\xi_1^D\partial^A\xi_{2D}-\xi_2^D\partial^A\xi_{1D}\bigg)
-\frac{1}{2}\partial^A\bigg(\xi_{2D}Z^{D}{}_E\xi_1^E\bigg),
\nn\\
\eea
where
\bea
Z\equiv Z^{A}{}_B\equiv\begin{pmatrix} -1 & 0
 \\ 0 & 1  \end{pmatrix}.
\eea
We use $\eta$ to raise or lower index for $Z$.
It is easy to find
\bea
\lbrack\delta_1, \delta_2\rbrack=-\delta_{\lbrack\xi_1, \xi_2\rbrack_F}.
\eea
We perform the B-transformation on the $C$-bracket and $F$-bracket with the strong constraints.
The B-transformation is defined by
\bea
e^B\equiv\begin{pmatrix} 1& 0\\ B& 1 \end{pmatrix}, \qquad e^B\begin{pmatrix} X\\ \xi \end{pmatrix}=\begin{pmatrix} X\\ \xi+BX \end{pmatrix}=\begin{pmatrix} X\\ \xi+i_XB \end{pmatrix}.
\eea
This transformation is a symmetry of the sigma model.
We show the calculation on the Courant bracket.
\bea
\lbrack e^B(X+\xi), e^B(Y+\eta)\rbrack_{\mbox{Cour}}&=&\lbrack X+\xi+i_X B, Y+\eta+i_Y B\rbrack_{\mbox{Cour}}
\nn\\
&=&\lbrack X+\xi, Y+\eta\rbrack_{\mbox{Cour}}+\lbrack X, i_Y B\rbrack_{\mbox{Cour}}+\lbrack i_X B, Y\rbrack_{\mbox{Cour}}
\nn\\
&=&\lbrack X+\xi, Y+\eta\rbrack_{\mbox{Cour}}+{\cal L}_X i_Y B-\frac{1}{2}di_Xi_Y B-{\cal L}_Yi_X B+\frac{1}{2}di_Yi_X B
\nn\\
&=&\lbrack X+\xi, Y+\eta\rbrack_{\mbox{Cour}} +i_{\lbrack X, Y\rbrack}B+i_Yi_XdB
\nn\\
&=&e^B\bigg(\lbrack X+\xi, Y+\eta\rbrack_{\mbox{Cour}}\bigg)+i_Yi_XdB.
\eea
If $dB=0$, we can obtain  automorphism. It shows that the symmetry of a theory governed by the Courant bracket can define a non-zero $H$-flux ($dH=0$) and possibly be extended to the $O(D, D)~ description$. For the closed string theory, we use the $O(D, D)$ structure to rewrite this theory. For the D-brane theory without the one-form gauge field, we should have the same story. Before we calculate the $F$-bracket, we define the notation for the $F$-bracket with the strong constraints
\bea
\lbrack X+\xi, Y+\eta\rbrack_{F}=\lbrack X, Y\rbrack+{\cal L}_{X}\eta-{\cal L}_{Y}\xi.
\eea
Therefore, we obtain
\bea
\lbrack e^B(X+\xi), e^B(Y+\eta)\rbrack_F&=&\lbrack X+\xi+i_X B, Y+\eta+i_Y B\rbrack_F
\nn\\
&=&\lbrack X+\xi, Y+\eta\rbrack_F+\lbrack X, i_Y B\rbrack_F+\lbrack i_X B, Y\rbrack_F
\nn\\
&=&\lbrack X+\xi, Y+\eta\rbrack_F+{\cal L}_X i_Y B-{\cal L}_Yi_X B
\nn\\
&=&\lbrack X+\xi, Y+\eta\rbrack_F +i_{\lbrack X, Y\rbrack}B+i_Yi_X dB-di_Yi_X B
\nn\\
&=&e^B\bigg(\lbrack X+\xi, Y+\eta\rbrack_F\bigg)+i_Yi_X dB-di_Yi_X B.
\eea
This means that we cannot use $dB=0$ to show automorphism as the Courant bracket. The information of the $F$-bracket shows that the $O(D, D)$ structure is not suitable to describe the DBI term.

\subsection{Classical Equivalence}
We prove classical equivalence between the double and normal sigma model in this section. We start from the bulk action
\bea
\label{bulk}
S_{\mbox{bulk}}=\frac{1}{2}\int d^2\sigma\ \bigg(\partial_1X^A{\cal H}_{AB}\partial_1X^B-\partial_1X^A\eta_{AB}\partial_0X^B\bigg).
\eea
The worldsheet metric is $(-, +)$ signature on the bulk. The equation of motion of $X^A$ is
\bea
\partial_1\bigg({\cal H}_{AB}\partial_1X^B-\eta_{AB}\partial_0X^B\bigg)=\frac{1}{2}\partial_1 X^B\partial_A{\cal H}_{BC}\partial_1X^C.
\eea
We use the strong constraints ($\tilde{\partial}^m$=0) to show the equivalence. Then we use the self-duality relation
\bea
\label{con1}
{\cal H}^m{}_B\partial_1X^B-\eta^m{}_B\partial_0X^B=0
\eea
to remove half degrees of freedom. It is equivalent to
\bea
\partial_1\tilde{X}=B\partial_1 X+g\partial_0 X.
\eea
The gauge transformation of $X^A$ is governed by the generalized Lie derivative as the generalized metric. The gauge transformation is
\bea
\delta X^A=\xi^C\partial_C X^A+(\partial^A\xi_C-\partial_C\xi^A)X^C.
\eea
We assume that the gauge parameters do not depend on the worldsheet coordinates. Then we can show (\ref{con1}) is covariant under the gauge transformation with $\tilde{\partial}^m$=0. It implies that  (\ref{con1}) do not need to be modified from the covariant property.
Then we substitute (\ref{con1}) to the other one equation of motion. 
\bea
\label{cl1}
&&\partial_1\bigg({\cal H}_{mB}\partial_1X^B-\eta_{mB}\partial_0X^B\bigg)
\nn\\
&=&\partial_1(g\partial_1 X+B\partial_0 X)_m-\partial_0(g\partial_0X+B\partial_1X)_m.
\eea
\bea
\label{cl2}
&&\frac{1}{2}\partial_1 X^B\partial_m{\cal H}_{BC}\partial_1X^C
\nn\\
&=&-\frac{1}{2}\partial_0X^p\partial_mg_{pq}\partial_0X^q+\frac{1}{2}\partial_1X^p\partial_mg_{pq}\partial_1X^q+\partial_1X^p\partial_mB_{pq}\partial_0X^q.
\eea
We combine (\ref{cl1}) and (\ref{cl2}) to find the same equations of motion as 
\bea
\frac{1}{2}\int d^2\sigma\ \bigg(\partial_{\alpha}X^m g_{mn}\partial^{\alpha}X^n-\epsilon^{\alpha\beta}\partial_{\alpha}X^mB_{mn}\partial_{\beta}X^n\bigg),
\eea
where $\alpha$=0, 1 (We indicate the worldsheet directions by the Greek index.).
If we impose the Neumann boundary condition on the $\sigma^1$ direction, the suitable boundary term for the double sigma model should be
\bea
S_{\mbox{boundary}}=-\int d\sigma^0\ A_m\partial_0X^m
\eea 
to guarantee the gauge invariance. This boundary term breaks the $O(D, D)$ structure with the consistent understanding from the $F$-bracket.

\section{One-Loop $\beta$ Function}
\label{3}
We implement the self-duality relation (\ref{con1}) at the off-shell level. Then we obtain the DBI theory from the one-loop $\beta$ function.

\subsection{Self-Duality Relation at the Off-Shell Level}
We can have the classical equivalence with the on-shell self-duality relation. But quantum fluctuation of the double sigma model needs the self-duality relation beyond the on-shell level. For the constant background fields, we can show it. We first set $B$=0. We can always redefine the one-form gauge field to find the non-zero constant $B$ field without losing generality. 
The equations of motion on the bulk is
\bea
\partial_1\big(\partial_1\tilde{X}-g\partial_0 X\big)&=&0,
\nn\\
\partial_1\big(g\partial_1X-\partial_0\tilde{X}\big)&=&0.
\eea
They can be rewritten as
\bea
\partial_1\tilde{X}-g\partial_0X&=&f,
\nn\\
\partial_1\big(g\partial_1X\big)-\partial_0\big(g\partial_0X\big)&=&\partial_0 f,
\eea
where $f$ is an arbitrary function of $\sigma^0$. Then we can redefine $X^m$ ($X^{m}\rightarrow X^{m}+h^{m}(\sigma^0)$) to find the consistent equations of motion with the self-duality relation. We assume $-g\partial_0 h=f$. Then we obtain
\bea
\partial_1\tilde{X}-g\partial_0X&=&0,
\nn\\
\partial_1\big(g\partial_1X\big)-\partial_0\big(g\partial_0X\big)&=&0.
\eea
The first equation is the self-duality relation and second equation is the equation of motion on the bulk. Then we discuss the equation of motion on the boundary. The Neumann boundary condition is 
\bea
g\partial_1X=F(X)\partial_0X.
\eea
This boundary condition is still an invariant form from the redefinition. The above discussion shows that we can have the self-duality relation at the off-shell level to describe the same equations of motion with the normal sigma model in the case of the constant background. The difficulty of quantization for the non-constant background is the same as the chiral boson theory \cite{Bernstein:1988zd, Imbimbo:1987yt, Labastida:1987zy, Floreanini:1987as}. Nevertheless, we will show that we can obtain the DBI theory from the one-loop $\beta$ function in the next section. 

\subsection{One-Loop $\beta$ Function}
We set $B=0$ and $g=I$ ($I\equiv$ identity matrix) to simplify the calculation without losing generality in the end of this section. 
We first show the standard calculation of the one-loop $\beta$ function as \cite{Abouelsaood:1986gd}. From the variation ($X\rightarrow X+\xi$) of 
the boundary term, we obtain 
\bea
-\int d\sigma^0\ \bigg(A_m\partial_0X^m+\xi^m F_{mn}\partial_0X^n+\frac{1}{2}\big(\xi^m\xi^n\partial_m F_{np}\partial_0X^p+\xi^m\partial_0
\xi^nF_{mn}\big)\bigg),
\nn\\
\eea
where
\bea
F_{mn}\equiv\partial_m A_n-\partial_n A_m. \qquad
\eea
Then the Green's function on the bulk are
\bea
\bigg({\cal H}^m{}_B\partial_1^2-\eta^m{}_B\partial_0\partial_1\bigg)G^{Bp}(\sigma, \sigma^{\prime})&=&iI^{mp}\delta^2(\sigma-\sigma^{\prime}),
\nn\\
\bigg({\cal H}_{mB}\partial_1^2-\eta_{mB}\partial_0\partial_1\bigg)G^B{}_p(\sigma, \sigma^{\prime})&=&iI_{mp}\delta^2(\sigma-\sigma^{\prime})
\eea
and on the boundary is 
\bea
{\cal H}_{mB}\partial_1G^{Bp}-F_{mn}\partial_0G^{np}=0.
\eea
The counter term is
\bea
-\frac{1}{2}\int d\sigma^0\ \Gamma_m\partial_0X^m,
\eea
where
\bea
\Gamma_m=\lim_{\epsilon\rightarrow 0}G^{np}(\epsilon\equiv\sigma^0-\sigma^{0\prime})\partial_nF_{pm}.
\eea
The $\beta$ function is 
\bea
\beta_m\equiv\epsilon\frac{\partial\Gamma_m}{\partial\epsilon}.
\eea
Then we try to solve the Green's functions on the bulk. We first change the coordinates
\bea
z=\sigma+\tau, \qquad \bar{z}=\sigma-\tau.
\eea
We only need to solve 
\bea
I^{mn}\bigg(\frac{1}{2}\partial_z^2+\frac{1}{2}\partial_{\bar{z}}^2+\partial_z\partial_{\bar{z}}\bigg)G_n{}^p&=&iI^{mp}\delta^2(z-z^{\prime}),
\nn\\
I_{mn}\bigg(\frac{1}{2}\partial_{z}^2+\frac{1}{2}\partial_{\bar{z}}^2+\partial_z\partial_{\bar{z}}\bigg)G^n{}_{p}&=&iI_{mp}\delta^2(z-z^{\prime})
\eea
with 
\bea
\delta^2(z-z^{\prime})\equiv\frac{1}{2}\delta^2(\sigma-\sigma^{\prime})
\eea
on the bulk.
The solutions of the Green's function on the bulk are
\bea
G_n{}^p&=&-\frac{I_n{}^p}{2\pi}\ln (\bar{z}-\bar{z}^{\prime})-\frac{I_n{}^p}{2\pi}\ln(z-z^{\prime}),
\nn\\
G^n{}_{p}&=&-\frac{I^n{}_p}{2\pi}\ln (\bar{z}-\bar{z}^{\prime})-\frac{I^n{}_{p}}{2\pi}\ln(z-z^{\prime}).
\eea
The equation of the Green's function on the boundary is followed by
\bea
&&{\cal H}_{mB}(\partial_z+\partial_{\bar{z}})G^{Bp}-F_{mn}(\partial_z-\partial_{\bar{z}})G^{np}
\nn\\
&=&(I_{mn}-
F_{mn})\partial_zG^{np}+(I_{mn}+F_{mn})\partial_{\bar{z}}G^{np}=0.
\eea
The solution of the Green's function on the boundary is
\bea
G^{np}&=&I^{np}\ln\mid z-z^{\prime}\mid+\frac{1}{2}(I-F)^{nq}(I+F)_{qw}I^{wp}\ln(z+\bar{z}^{\prime})
\nn\\
&&+\frac{1}{2}(I+F)^{nq}(I-F)_{qw}I^{wp}\ln(\bar{z}+z^{\prime})\mid_{z=-\bar{z},\  z^{\prime}=-\bar{z}^{\prime}}.
\eea
Therefore, the $\beta$ function is 
\bea
\beta_m=\bigg(I^{np}+\frac{1}{2}(I-F)^{nq}(I+F)_{qw}I^{wp}+\frac{1}{2}(I+F)^{nq}(I-F)_{qw}I^{wp}\bigg)\partial_n F_{pm}.
\nn\\
\eea
We rewrite it by
\bea
&&I+\frac{1}{2}\bigg((I-F)^{-1}(I+F)+(I+F)^{-1}(I-F)\bigg)
\nn\\
&=&I+\frac{1}{2}(I-F^2)^{-1}\bigg((I+F^2)+(I+F^2)\bigg)=I+(I-F^2)^{-1}(F^2+I)
\nn\\
&=&2(I-F^2)^{-1}.
\eea
So that we can get
\bea
\beta_m=2(I-F^2)^{np}\partial_nF_{pm},
\eea
where
\bea
(I-F^2)^{np}\equiv(I-F^2)^{-1}.
\eea
Let us show an useful identity 
\bea
(I-F^2)^{mn}\beta_n=-2\bigg\lbrack\partial^p\bigg((I-F^2)^{mn}F_{np}\bigg)-(I-F^2)^{mx}F_{xr}\partial^nF^r{}_q(I-F^2)^{qp}F_{pn}\bigg\rbrack
\nn\\
\eea
by
\bea
(a+b)^{-1}&=&a^{-1}-a^{-1}(I+ba^{-1})^{-1}ba^{-1},
\nn\\
(I-F^2)^{-1}&=&I+(I-F^2)^{-1}F^2.
\eea
We can use the Bianchi identity to rewrite it.
\bea
&&\frac{1}{4}\bigg(\frac{F}{I-F^2}\bigg)_{mn}\partial^n\mbox{Tr}\ln(I-F^2)
\nn\\
&=&-\frac{1}{4}\bigg(\frac{F}{I-F^2}\bigg)_{mn}\frac{1}{I-F^2}\partial^n(F^2)
\nn\\
&=&-\frac{1}{2}\bigg(\frac{F}{I-F^2}\bigg)_{mn}(I-F^2)^{pq}F_{pr}\partial^nF^r{}_q
\nn\\
&=&-\frac{1}{2}\bigg(\frac{F}{I-F^2}\bigg)_{mn}\partial^nF^{pq}\bigg(\frac{F}{I-F^2}\bigg)_{qp}
\nn\\
&=&-\frac{1}{2}\bigg(\frac{F}{I-F^2}\bigg)_{mn}\bigg(-\partial^pF^{qn}-\partial^qF^{np}
\bigg)\bigg(\frac{F}{I-F^2}\bigg)_{qp}
\nn\\
&=&\bigg(\frac{F}{I-F^2}\bigg)_{mn}(\partial^pF^{qn})\bigg(\frac{F}{I-F^2}\bigg)_{qp}
\nn\\
&=&-\bigg(\frac{F}{I-F^2}\bigg)_{mn}\partial^pF^{nq}\bigg(\frac{F}{I-F^2}\bigg)_{qp},
\eea
where
\bea
(I-F^2)^{-1}F\equiv\bigg(\frac{F}{I-F^2}\bigg).
\eea
After that, we can obtain 
\bea
(I-F^2)^{mn}\beta_n=-2\bigg\lbrack\partial^p\bigg((I-F^2)^{mn}F_{np}\bigg)+\frac{1}{4}\bigg(\frac{F}{I-F^2}\bigg)_{mn}\partial^{n}\mbox{Tr}\ln(I-F^2)\bigg\rbrack.
\nn\\
\eea
By multiplying $\sqrt{\det(I+F)}$, the equation of motion of the DBI action can be rewritten as
\bea
\sqrt{\det(I+F)}(I-F^2)^{mn}\beta_n=0.
\eea
It is equivalent to $\beta_m=0$. We consistently obtain
\bea
\sqrt{\det\bigg(I+F\bigg)}.
\eea
Then we show how to obtain the effective action for the general constant metric $g$.
Because $g$ is a symmetric matrix, we can diagonalize $g$. Then we rescale the diagonal matrix and redefine the one-form gauge field.
We equivalently obtain
\bea
\sqrt{\det\bigg(g+F\bigg)}.
\eea
This calculation shows that this double sigma model can be a consistent model with quantum fluctuation. It is a non-trivial consistent check beyond the classical equivalence.

\section{The Generalized Metric Formulation}
\label{4}
We construct the low energy effective action based on the symmetry point of view. This action is written in terms of the generalized metric and scalar dilaton. We use the equations of motion to define the generalized scalar curvature and tensor.
\subsection{The Low Energy Effective Action}
We construct this low energy effective action in two parts. The first part is based on the diffeomorphism and one-form gauge transformation. The candidate is the DBI action. The second part is based on the  $O(D, D)$ structure, $\mathbb{Z}_2$ symmetry, gauge symmetry with strong constraints and two derivative terms. We first discuss the $\mathbb{Z}_2$ symmetry
\bea
B_{mn}\rightarrow -B_{mn},
\qquad
\tilde{\partial}^m\rightarrow -\tilde{\partial}^m.
\eea
For $\tilde{\partial}^m\rightarrow -\tilde{\partial}^m$, we can rewrite it as
\be
\partial_A\rightarrow  Z\, \partial_A \,.
\ee
The transformation of ${\cal H}^{AB}$ under the transformation $B_{mn} \rightarrow - B_{mn}$ can be written as
\be
{\cal H}^{AB}  \to  Z {\cal H}^{AB} Z\,, \qquad
{\cal H}_{AB}  \to  Z {\cal H}_{AB} Z \,.
\ee
Then the action can be constructed with respect to the gauge symmetry (with the strong constraints) from all possible $O(D, D)$ elements ($\partial_A$, ${\cal H}^{AB}$, ${\cal H}_{AB}$ and $d$) up to a boundary term. The theory is
\bea
\label{acg}
S_2 &=& \int dx \ d\tilde x  \
   e^{-2d}\Big(\frac{1}{8}{\cal H}^{AB}\partial_{A}{\cal H}^{CD}
  \partial_{B}{\cal H}_{CD}-\frac{1}{2}
  \,{\cal H}^{AB}\partial_{B}{\cal H}^{CD}\partial_{D}
  {\cal H}_{AC}
\nn\\
  &&-2\partial_{A}d\partial_{B}{\cal H}^{AB}+4{\cal H}^{AB}\,\partial_{A}d
  \partial_{B}d \Big).
 \eea
 This action is uniquely determined from the above criteria. For the goal of rewriting total theory without using the field strength or one-form gauge field, we redefine the generalized metric by
\bea
B_{mn}\rightarrow B_{mn}-F_{mn}.
\eea
This field redefinition does not modify all results of the closed string part.
The action of the DBI part is 
\bea
S_1 &=& \int dx \ d\tilde x  \ e^{-d}\bigg(-\det( {\cal H}_{mn} )\bigg)^{\frac{1}{4}}.
\eea
We only rewrite the DBI part from ${\cal H}_{mn}$ and $d$ because the boundary conditions are not modified from the strong constraints except for the fields do not depend on the dual coordinates.
The $e^{-d}$ shows that the T-duality changes the dimensions of spacetime. It reflects the difference between the closed and open string. This dilaton term and ${\cal H}_{mn}=\bigg(g-(B-F)g^{-1}(B-F)\bigg)_{mn}$  give the manifest equivalence between the closed and open string parameters. Because the manifest equivalence of the double field theory does not change spacetime dimensions, the suitable manifest equivalence should not be the T-duality. This manifest equivalence without the field strength can be the manifest Buscher's rule. We can say that this DBI action has the manifest T-duality rule without the field strength as the massless closed string theory. Nevertheless, it is not the T-duality because the T-duality for the open string should change spacetime dimensions. 
The total action is
\bea
S&=&S_1+\alpha S_2
\nn\\
&=&\int dx \ d\tilde x  \ e^{-d}\bigg[\bigg(-\det( {\cal H}_{mn} )\bigg)^{\frac{1}{4}}\bigg]
\nn\\
&&+\alpha e^{-2d}\bigg[\bigg(\frac{1}{8}{\cal H}^{AB}\partial_{A}{\cal H}^{CD}
  \partial_{B}{\cal H}_{CD}-\frac{1}{2}
  \,{\cal H}^{AB}\partial_{B}{\cal H}^{CD}\partial_{D}
  {\cal H}_{AC}
\nn\\
  &&-2\partial_{A}d\partial_{B}{\cal H}^{AB}+4{\cal H}^{AB}\,\partial_{A}d
  \partial_{B}d\bigg) \Bigg],
\eea
where $\alpha$ is an arbitrary constant.
If we use $\tilde{\partial}^m$=0, we obtain
\bea
\int dx\ \sqrt{-\det g}\bigg[e^{-\phi}\bigg( -\det(g+B-F)\bigg)^{\frac{1}{2}}\bigg( -\det g\bigg)^{-\frac{1}{2}}+\alpha e^{-2\phi}\bigg(R+4(\partial\phi)^2-\frac{1}{12}H^2\bigg)\bigg],
\nn\\
\eea
where $R$ is the Ricci scalar and $H=dB$ is the three form field strength. This theory is determined from the symmetry point of view up to a relative coefficient. This coefficient can be determined from the one-loop $\beta$ function. This action is also consistent with \cite{Callan:1986bc}. 
If we set $D$=10, it is the low energy effective theory of the D9-brane on the curved background. Then the nontrivial flux can be realized on the D-brane theory. After we perform the T-duality on this theory, we should find the non-geometric flux on the lower dimensions. 

\subsection{Generalized Scalar Curvature and Tensor}
We show the equations of the motion in this section. We also define the generalized scalar curvature and tensor from the equations of motion.
We first define the equations of motion for $d$ to be the generalized scalar curvature.
\bea
{\cal R}&\equiv&\frac{1}{2}
\bigg(-\det( {\cal H}_{mn} )\bigg)^{\frac{1}{4}}
\nn\\
 &&+\alpha\bigg(4{\cal H}^{AB}\partial_A\partial_B d-\partial_A\partial_B{\cal H}^{AB}-4{\cal H}^{AB}\partial_A d\partial_B d+4\partial_A{\cal H}^{AB}\partial_B d
\nn\\
&&+\frac{1}{8}{\cal H}^{AB}\partial_A{\cal H}^{CD}\partial_B{\cal H}_{CD}
-\frac{1}{2}{\cal H}^{AB}\partial_A{\cal H}^{CD}\partial_C{\cal H}_{BD}\bigg).
\eea
It satisfies the suitable symmetry
\bea
\delta_{\xi}{\cal R}=\xi^A\partial_A{\cal R}
\eea
with $\tilde{\partial}^m$=0 for the closed string part. Considering the DBI part, we do not have diffeomorphism. It shows the difference between the closed and open string theory.

At the end of this section, we vary the $\cal{H}^{CD}$, which provides the generalized Ricci tensor. To calculate the variation, we need to introduce one auxiliary field in the action
\bea
e^{-2d}\lambda_{mn}({\cal H}\eta{\cal H}-\eta)^{mn}.
\eea
From the variation of ${\cal H}^{CD}$, we can get
\bea
{\cal K}_{CD}+(\lambda S+S^{t}\lambda)_{CD}=0,
\eea
where
\bea
&&\lambda\equiv\lambda_{\bullet\bullet},
\qquad S\equiv{\cal H}\eta,
\qquad
S^2=1,
\nn\\
&&{\cal K}_{AB}\equiv e^{d}\frac{\delta\bigg(-\det({\cal H}_{pq})\bigg)^{\frac{1}{4}}}{\delta{\cal H}_{AB}}
\nn\\
&&\ \ \ \ \ \ \ \ \ \ +\alpha\bigg(\frac{1}{8}\partial_A{\cal H}^{CD}\partial_B{\cal H}_{CD}-\frac{1}{4}(\partial_D-2(\partial_D d))({\cal H}^{DC}\partial_C{\cal H}_{AB})+2\partial_A\partial_B d
\nn\\
&&\ \ \ \ \ \ \ \ \ \  -\frac{1}{2}\partial_{(A}{\cal H}^{CD}\partial_D{\cal H}_{B)C}+\frac{1}{2}(\partial_D-2(\partial_D d))({\cal H}^{CD}\partial_{(A}{\cal H}_{B)C}+{\cal H}^C{}_{(A}\partial_C{\cal H}^D{}_{B)})\bigg)
\nn\\
\eea
except for 
\bea
\frac{\delta\bigg(-\det({\cal H}_{pq})\bigg)^{\frac{1}{4}}}{\delta{\cal H}_{mn}}=\frac{1}{4}\bigg(-\det({\cal H}_{pq})\bigg)^{\frac{1}{4}}\bigg(({\cal H}_{pq})^{-1}\bigg)^{mn}.
\eea
Other variation of the generalized metric do not give non-zero contribution.
The equation of motion of ${\cal H}^{CD}$ is equivalent to
\bea
(S^{t}{\cal K}S)_{CD}+(S^{t}\lambda+\lambda S)_{CD}=0.
\eea
It implies
\bea
{\cal K}_{CD}-(S^{t}{\cal K}S)_{CD}=0.
\eea
Therefore, we define the generalized Ricci tensor 
\bea
{\cal R}_{CD}\equiv \frac{1}{2}{\cal K}_{CD}-\frac{1}{2}(S^{t}{\cal K}S)_{CD}.
\eea
We provide the generalized scalar curvature and generalized tensor from the equations of motion.


\section{Conclusion}
\label{5}
We compute the one-loop $\beta$ function for the double sigma model with the strong constraints in the constant background. We can obtain the consistent low energy effective theory, the DBI theory. It shows that the construction of this double sigma model is calculable for quantum corrections. A calculable double sigma model as normal sigma model is important to understand new physics. Although this calculation is only in the case of the constant background, it is still an important step for double field theory. So far we did not have any consistent check with quantum fluctuation for double sigma model of open string. We also rewrite the low energy effective theory in terms of the generalized metric and scalar dilaton. This construction also leads us to define the generalized scalar curvature and tensor. It is the usefulness of the generalized metric formulation \cite{Hohm:2010pp}. 

This double sigma model provides a general way to put the boundary term. Not only restricted to the open string sigma model, it should be able to extend to the other sigma model. Double sigma model of open string provides a possibility to unify all string theory in the language of double field theory. It should be used in all different kind of theories, not only for some special theories. We believe that this work opens a new door to reformulate these theories by a more powerful formulation for the T-fold problem. 

The formulation of open string relies on the boundary conditions. Choosing boundary conditions is equivalent to choosing the boundary terms. It should be interesting to embed different boundary conditions in the projectors. From choosing projectors to determine the boundary conditions should be interesting. Then the other one interesting issue is to use canonical way to quantize this open string theory to find the non-commutative relation. We leave boundary conditions and quantization two interesting problems to the future.

The most serious problem of the double field theory is to relax strong constraints to obtain more physical solutions. The difficulty is that the generalized Lie derivative is not closed without the strong constraints. The way is to develop new algebraic structures or introduce more fields. However, we do not get more understanding from the open string than the closed string. We still need to go back to the closed string to understand this problem.

The D-brane theory can be lifted to the M5-brane theory. The construction of the D-brane theory should shed the light on understanding properties of the M5-brane theory. In this low energy D-brane theory, we can find the non-geometric flux from the T-duality. At the low energy level, we should also find the non-geometric flux on the M5-brane. It should be interesting to study it.

From symmetry point of view, we can deduce the low energy effective action up to a relative coefficient. It should be interesting to obtain this coefficient from symmetry between the closed and open string theories without calculating one-loop $\beta$ function. The answer may be hidden in the $\alpha^{\prime}$ correction of the closed string theory. However, the probe of principles for brane theory is an interesting direction. It should help us to find the action for the M5-brane theory from these principles.

Before we worked on the non-geometric frame to study the non-geometric flux in the massless closed string theory, we can see that this non-geometric frame in the open string theory is equivalent to using the open string parameters. From this manifest formulation, it should offer a clear picture. We do not know how to deal with the non-geometric problems on the commutative space, but the extension of the non-geometric flux can be defined on the non-commutative space or by the open string parameters. The non-commutative space possibly be a more natural space to describe string theory than the commutative space.

\acknowledgments

The author would like to thank David S. Berman, Chong--Sun Chu, Jun-Kai Ho and Xing Huang for the useful discussion.  This work is supported in part by CASTS (grant \#103R891003), Taiwan, R.O.C..


\begin{thebibliography}{99}

\bibitem{Zwiebach:1992ie} 
  B.~Zwiebach,
  \emph{Closed string field theory: Quantum action and the B-V master equation},
  \emph{Nucl.\ Phys.\ B} {\bf 390} (1993) 33
  [hep-th/9206084].

\bibitem{Saadi:1989tb} 
  M.~Saadi and B.~Zwiebach,
  \emph{Closed String Field Theory from Polyhedra},
  \emph{Annals Phys.\ } {\bf 192} (1989) 213.
   
\bibitem{Ho:2008nn}
  P.~-M.~Ho and Y.~Matsuo,
  \emph{M5 from M2},
  \emph{JHEP} {\bf 0806} (2008) 105
  [arXiv:0804.3629 [hep-th]].

\bibitem{Ho:2012dn}
  P.~-M.~Ho, C.~-T.~Ma and C.~-H.~Yeh,
  \emph{BPS States on M5-brane in Large C-field Background},
  \emph{JHEP} {\bf 1208} (2012) 076
  [arXiv:1206.1467 [hep-th]].

\bibitem{Ho:2013opa} 
  P.~M.~Ho and C.~T.~Ma,
  \emph{S-Duality for D3-Brane in NS-NS and R-R Backgrounds},
  \emph{JHEP} {\bf 1411} (2014) 142
  [arXiv:1311.3393 [hep-th]].

\bibitem{Ho:2013paa} 
  P.~M.~Ho and C.~T.~Ma,
  \emph{Effective Action for Dp-Brane in Large RR (p-1)-Form Background},
  \emph{JHEP} {\bf 1305} (2013) 056
  [arXiv:1302.6919 [hep-th]].

\bibitem{Ma:2012hc} 
  C.~T.~Ma and C.~H.~Yeh,
  \emph{Supersymmetry and BPS States on D4-brane in Large C-field Background},
  \emph{JHEP} {\bf 1303} (2013) 131
  [arXiv:1210.4191 [hep-th]].

\bibitem{Jurco:2012yv}
  B.~Jurco and P.~Schupp,
  \emph{Nambu-Sigma model and effective membrane actions},
  \emph{Phys.\ Lett.\ B} {\bf 713} (2012) 313
  [arXiv:1203.2910 [hep-th]].

\bibitem{Schupp:2012nq} 
  P.~Schupp and B.~Jurco,
  \emph{Nambu Sigma Model and Branes},
  \emph{PoS CORFU} {\bf 2011} (2011) 045
  [arXiv:1205.2595 [hep-th]].

\bibitem{Ho:2014una} 
  J.~K.~Ho and C.~T.~Ma,
  \emph{Dimensional Reduction of the Generalized DBI},
  arXiv:1410.0972 [hep-th].
 
\bibitem{Jurco:2013upa} 
  B.~Jurco, P.~Schupp and J.~Vysoky,
  \emph{On the Generalized Geometry Origin of Noncommutative Gauge Theory},
  \emph{JHEP} {\bf 1307} (2013) 126
  [arXiv:1303.6096 [hep-th]].

\bibitem{Jurco:2014sda} 
  B.~Jurčo, P.~Schupp and J.~Vysoký,
  \emph{Extended generalized geometry and a DBI-type effective action for branes ending on branes},
  \emph{JHEP} {\bf 1408} (2014) 170
  [arXiv:1404.2795 [hep-th]].

\bibitem{Lee:2010ey} 
  K.~Lee and J.~H.~Park,
  \emph{Partonic description of a supersymmetric p-brane},
  \emph{JHEP} {\bf 1004} (2010) 043
  [arXiv:1001.4532 [hep-th]].

\bibitem{Park:2008qe} 
  J.~H.~Park and C.~Sochichiu,
  \emph{Taking off the square root of Nambu-Goto action and obtaining Filippov-Lie algebra gauge theory action},
  \emph{Eur.\ Phys.\ J.\ C} {\bf 64} (2009) 161
  [arXiv:0806.0335 [hep-th]].

\bibitem{Hohm:2011dv}
  O.~Hohm, S.~K.~Kwak and B.~Zwiebach,
  \emph{Double Field Theory of Type II Strings},  \emph{JHEP} {\bf 1109} (2011) 013  [arXiv:1107.0008 [hep-th]].  

\bibitem{Hohm:2010xe}
  O.~Hohm and S.~K.~Kwak,
  \emph{Frame-like Geometry of Double Field Theory},  \emph{J.\ Phys.\ A} {\bf 44} (2011) 085404  [arXiv:1011.4101 [hep-th]].  

\bibitem{Hull:2009mi}
  C.~Hull and B.~Zwiebach,
  \emph{Double Field Theory},  \emph{JHEP} {\bf 0909} (2009) 099  [arXiv:0904.4664 [hep-th]].  

\bibitem{Hohm:2010jy}
  O.~Hohm, C.~Hull and B.~Zwiebach,
  \emph{Background independent action for double field theory},  \emph{JHEP} {\bf 1007} (2010) 016  [arXiv:1003.5027 [hep-th]].  

\bibitem{Lee:2013hma} 
  K.~Lee and J.~H.~Park,
  \emph{Covariant action for a string in "doubled yet gauged" spacetime},
  \emph{Nucl.\ Phys.\ B} {\bf 880} (2014) 134
  [arXiv:1307.8377 [hep-th]].

\bibitem{Tseytlin:1990va} 
  A.~A.~Tseytlin,
  \emph{Duality symmetric closed string theory and interacting chiral scalars},
  \emph{Nucl.\ Phys.\ B} {\bf 350} (1991) 395.

\bibitem{Tseytlin:1990nb} 
  A.~A.~Tseytlin,
  \emph{Duality Symmetric Formulation of String World Sheet Dynamics},
  \emph{Phys.\ Lett.\ B} {\bf 242} (1990) 163.

\bibitem{Copland:2011wx} 
  N.~B.~Copland,
  \emph{A Double Sigma Model for Double Field Theory},
  \emph{JHEP} {\bf 1204} (2012) 044
  [arXiv:1111.1828 [hep-th]].

\bibitem{Berman:2007yf} 
  D.~S.~Berman and D.~C.~Thompson,
  \emph{Duality Symmetric Strings, Dilatons and O(d,d) Effective Actions},
  \emph{Phys.\ Lett.\ B} {\bf 662} (2008) 279
  [arXiv:0712.1121 [hep-th]].

\bibitem{Berman:2007xn} 
  D.~S.~Berman, N.~B.~Copland and D.~C.~Thompson,
  \emph{Background Field Equations for the Duality Symmetric String},
  \emph{Nucl.\ Phys.\ B} {\bf 791} (2008) 175
  [arXiv:0708.2267 [hep-th]].

\bibitem{Avramis:2009xi} 
  S.~D.~Avramis, J.~P.~Derendinger and N.~Prezas,
  \emph{Conformal chiral boson models on twisted doubled tori and non-geometric string vacua},
  \emph{Nucl.\ Phys.\ B} {\bf 827} (2010) 281
  [arXiv:0910.0431 [hep-th]].

\bibitem{Duff:1989tf} 
  M.~J.~Duff,
  \emph{Duality Rotations in String Theory},
  \emph{Nucl.\ Phys.\ B} {\bf 335} (1990) 610.

\bibitem{Berkeley:2014nza} 
  J.~Berkeley, D.~S.~Berman and F.~J.~Rudolph,
  \emph{Strings and Branes are Waves},
  \emph{JHEP} {\bf 1406} (2014) 006
  [arXiv:1403.7198 [hep-th]].

\bibitem{Siegel:1993xq} 
  W.~Siegel,
  \emph{Two vierbein formalism for string inspired axionic gravity},
  \emph{Phys.\ Rev.\ D} {\bf 47} (1993) 5453 
  [hep-th/9302036].

\bibitem{Siegel:1993th} 
  W.~Siegel,
  \emph{Superspace duality in low-energy superstrings},
  \emph{Phys.\ Rev.\ D} {\bf 48} (1993) 2826
  [hep-th/9305073].

\bibitem{Siegel:1993bj} 
  W.~Siegel,
  \emph{Manifest duality in low-energy superstrings},
  [hep-th/9308133].

\bibitem{Hull:2009zb}
  C.~Hull and B.~Zwiebach,
  \emph{The Gauge algebra of double field theory and Courant brackets},  \emph{JHEP} {\bf 0909} (2009) 090  [arXiv:0908.1792 [hep-th]].  
  
\bibitem{Hohm:2010pp}
  O.~Hohm, C.~Hull and B.~Zwiebach,
  \emph{Generalized metric formulation of double field theory},  \emph{JHEP} {\bf 1008} (2010) 008  [arXiv:1006.4823 [hep-th]].  

\bibitem{Aldazabal:2011nj}
  G.~Aldazabal, W.~Baron, D.~Marques and C.~Nunez,
  \emph{The effective action of Double Field Theory},  \emph{JHEP} {\bf 1111} (2011) 052  [Erratum-ibid.\  {\bf 1111} (2011) 109 ]  [arXiv:1109.0290 [hep-th]].  

\bibitem{Andriot:2012an} 
  D.~Andriot, O.~Hohm, M.~Larfors, D.~Lust and P.~Patalong,
  \emph{Non-Geometric Fluxes in Supergravity and Double Field Theory},
  \emph{Fortsch.\ Phys.\ } {\bf 60} (2012) 1150 
  [arXiv:1204.1979 [hep-th]].
 
\bibitem{Kimura:2014upa} 
  T.~Kimura, S.~Sasaki and M.~Yata,
  \emph{World-volume Effective Actions of Exotic Five-branes},
  \emph{JHEP} {\bf 1407} (2014) 127
  [arXiv:1404.5442 [hep-th]].

\bibitem{Geissbuhler:2013uka} 
  D.~Geissbuhler, D.~Marques, C.~Nunez and V.~Penas,
  \emph{Exploring Double Field Theory},
  \emph{JHEP} {\bf 1306} (2013) 101
  [arXiv:1304.1472 [hep-th]].

\bibitem{Ma:2014ala} 
  C.~T.~Ma and C.~M.~Shen,
  \emph{Cosmological Implications from O(D,D)},
  \emph{Fortsch.\ Phys.\ }  {\bf 62} (2014) 921
  [arXiv:1405.4073 [hep-th]].

\bibitem{Hohm:2013jaa} 
  O.~Hohm, W.~Siegel and B.~Zwiebach,
  \emph{Doubled $\alpha'$-geometry},
  \emph{JHEP} {\bf 1402} (2014) 065
  [arXiv:1306.2970 [hep-th]].

\bibitem{Hohm:2013bwa}
  O.~Hohm, D.~Lust and B.~Zwiebach,
  \emph{The Spacetime of Double Field Theory: Review, Remarks, and Outlook},  \emph{Fortsch.\ Phys.\ } {\bf 61}, 926 (2013)  [arXiv:1309.2977 [hep-th]].  

\bibitem{Aldazabal:2013sca}
  G.~Aldazabal, D.~Marques and C.~Nunez,
  \emph{Double Field Theory: A Pedagogical Review},  \emph{Class.\ Quant.\ Grav.\ } {\bf 30} (2013) 163001  [arXiv:1305.1907 [hep-th]].  

\bibitem{Berman:2013eva} 
  D.~S.~Berman and D.~C.~Thompson,
  \emph{Duality Symmetric String and M-Theory},
  arXiv:1306.2643 [hep-th].

\bibitem{Hohm:2013vpa}
  O.~Hohm and H.~Samtleben,
  \emph{Exceptional Field Theory I: $E_{6(6)}$ covariant Form of M-Theory and Type IIB},
  \emph{Phys.\ Rev.\ D} {\bf 89} (2014) 066016
  [arXiv:1312.0614 [hep-th]].

\bibitem{Hohm:2013uia}
  O.~Hohm and H.~Samtleben,
  \emph{Exceptional Field Theory II: E$_{7(7)}$},
  \emph{Phys.\ Rev.\ D} {\bf 89} (2014) 066017
  [arXiv:1312.4542 [hep-th]].

\bibitem{Hohm:2014fxa} 
  O.~Hohm and H.~Samtleben,
  \emph{Exceptional Field Theory III: E$_{8(8)}$},
  \emph{Phys.\ Rev.\ D} {\bf 90} (2014) 066002
  [arXiv:1406.3348 [hep-th]].

\bibitem{Berman:2010is}
  D.~S.~Berman and M.~J.~Perry,
  \emph{Generalized Geometry and M theory},  \emph{JHEP} {\bf 1106} (2011) 074  [arXiv:1008.1763 [hep-th]].  

\bibitem{Hull:2004in} 
  C.~M.~Hull,
  \emph{A Geometry for non-geometric string backgrounds},
  \emph{JHEP} {\bf 0510} (2005) 065
  [hep-th/0406102].

\bibitem{Albertsson:2008gq}
  C.~Albertsson, T.~Kimura and R.~A.~Reid-Edwards,
  \emph{D-branes and doubled geometry},
  \emph{JHEP} {\bf 0904} (2009) 113
  [arXiv:0806.1783 [hep-th]].

\bibitem{Albertsson:2011ux}
  C.~Albertsson, S.~-H.~Dai, P.~-W.~Kao and F.~-L.~Lin,
  \emph{Double Field Theory for Double D-branes},
  \emph{JHEP} {\bf 1109} (2011) 025
  [arXiv:1107.0876 [hep-th]].

\bibitem{Gualtieri:2003dx}
  M.~Gualtieri,
  \emph{Generalized complex geometry},  math/0401221 [math-dg].  

\bibitem{Hitchin:2004ut}
  N.~Hitchin,
  \emph{Generalized Calabi-Yau manifolds},  \emph{Quart.\ J.\ Math.\ Oxford Ser.\ } {\bf 54} (2003) 281  [math/0209099 [math-dg]].  

\bibitem{Cavalcanti:2011wu} 
  G.~R.~Cavalcanti and M.~Gualtieri,
  \emph{Generalized complex geometry and T-duality},
  [arXiv:1106.1747 [math.DG]].

\bibitem{Hatsuda:2012uk} 
  M.~Hatsuda and T.~Kimura,
  \emph{Canonical approach to Courant brackets for D-branes},
  \emph{JHEP} {\bf 1206} (2012) 034
  [arXiv:1203.5499 [hep-th]].

\bibitem{Asakawa:2012px}
  T.~Asakawa, S.~Sasa and S.~Watamura,
  \emph{D-branes in Generalized Geometry and Dirac-Born-Infeld Action},
  \emph{JHEP} {\bf 1210} (2012) 064
  [arXiv:1206.6964 [hep-th]].

\bibitem{Ma:2014kia} 
  C.~T.~Ma,
  \emph{Gauge Transformation of Double Field Theory for Open String},
  arXiv:1411.0287 [hep-th].

\bibitem{Zwiebach:1985uq} 
  B.~Zwiebach,
  \emph{Curvature Squared Terms and String Theories},
  \emph{Phys.\ Lett.\ B} {\bf 156} (1985) 315.

\bibitem{Asakawa:2014kua} 
  T.~Asakawa, H.~Muraki, S.~Sasa and S.~Watamura,
  \emph{Poisson-generalized geometry and $R$-flux},
  arXiv:1408.2649 [hep-th].

\bibitem{Bernstein:1988zd} 
  M.~Bernstein and J.~Sonnenschein,
  \emph{A Comment on the Quantization of Chiral Bosons},
  \emph{Phys.\ Rev.\ Lett.\ } {\bf 60} (1988) 1772.

\bibitem{Imbimbo:1987yt} 
  C.~Imbimbo and A.~Schwimmer,
  \emph{The Lagrangian Formulation of Chiral Scalars},
  \emph{Phys.\ Lett.\ B} {\bf 193} (1987) 455.

\bibitem{Labastida:1987zy} 
  J.~M.~F.~Labastida and M.~Pernici,
  \emph{On the {BRST} Quantization of Chiral Bosons},
  \emph{Nucl.\ Phys.\ B} {\bf 297} (1988) 557.

\bibitem{Floreanini:1987as} 
  R.~Floreanini and R.~Jackiw,
  \emph{Selfdual Fields as Charge Density Solitons},
  \emph{Phys.\ Rev.\ Lett.\ } {\bf 59} (1987) 1873.
  
\bibitem{Abouelsaood:1986gd} 
  A.~Abouelsaood, C.~G.~Callan, Jr., C.~R.~Nappi and S.~A.~Yost,
  \emph{Open Strings in Background Gauge Fields},
  \emph{Nucl.\ Phys.\ B} {\bf 280} (1987) 599.

\bibitem{Callan:1986bc} 
  C.~G.~Callan, Jr., C.~Lovelace, C.~R.~Nappi and S.~A.~Yost,
  \emph{String Loop Corrections to beta Functions},
  \emph{Nucl.\ Phys.\ B} {\bf 288} (1987) 525.





\end{thebibliography}
\end{document}